\documentclass[reprint, 10pt, amsmath, amssymb, floatfix, aps, prd,  superscriptaddress, nofootinbib, nobibnotes,twocolumn]{revtex4-1}

\usepackage{graphicx}
\graphicspath{{figures/}}
\usepackage[utf8]{inputenc}
\usepackage{amsmath}
\usepackage{amsfonts}
\usepackage{amssymb}
\usepackage{multirow}
\usepackage{tcolorbox}
\usepackage{CJKutf8}
\usepackage{color}
\usepackage[colorlinks=true, linkcolor=red, citecolor=blue]{hyperref}
\usepackage{orcidlink}
\usepackage[capitalise]{cleveref}
\usepackage{acro}
\usepackage{adjustbox}
\usepackage{array}
\usepackage{amsmath}
\usepackage{natbib}
\usepackage{dblfnote}
\DFNalwaysdouble
\usepackage{slashed}
\usepackage{dcolumn}
\usepackage{hyperref}
\usepackage{geometry}
\usepackage{makecell}
\usepackage{subfig} 
\usepackage{overpic}
\usepackage{inputenc}
\usepackage{caption}
\def\({\left(}
\def\){\right)}
\def\[{\left[}
\def\]{\right]}
\def\be{\begin{eqnarray}}
\def\ee{\end{eqnarray}}

\DeclareAcronym{GW}{
  short = GW ,
  long = gravitational wave ,
  short-plural = s 
}
\DeclareAcronym{LIGO}{
  short = LIGO ,
  long = Laser Interferometer Gravitational-wave Observatory ,
  short-plural = 
}
\DeclareAcronym{LISA}{
  short = LISA ,
  long = Laser Interferometer Space Antenna ,
  short-plural =  
}
\DeclareAcronym{SKA}{
  short = SKA ,
  long = Square Kilometre Array ,
  short-plural =  
}  

\DeclareAcronym{PGW}{
  short = PGW ,
  long = primordial gravitational wave ,
  short-plural = s 
}
\DeclareAcronym{TSIGW}{
	short = TSIGW ,
	long = tensor-scalar induced gravitational wave ,
	short-plural =  s
}
  
\DeclareAcronym{SNR}{
	short = SNR ,
	long = signal-to-noise ratio ,
	short-plural = 
}

\DeclareAcronym{PTA}{
	short = PTA ,
	long = pulsar timing array ,
	short-plural = 
}

\DeclareAcronym{KDE}{
  short = KDE ,
  long = kernel density estimator ,
  short-plural = s
}

\DeclareAcronym{FLRW}{
  short = FLRW ,
  long = Friedmann-Lemaitre-Robertson-Walker ,
  short-plural =  
}

\DeclareAcronym{SIGW}{
	short = SIGW ,
	long = scalar induced gravitational wave ,
	short-plural =  s
}

\DeclareAcronym{TIGW}{
	short = TIGW ,
	long = tensor induced gravitational wave ,
	short-plural =  s
}

\DeclareAcronym{PBH}{
	short = PBH ,
	long = primordial black hole ,
	short-plural =  s
}

\DeclareAcronym{SMBHB}{
  short = SMBHB ,
  long = supermassive black hole binary ,
  short-plural = s
}

\DeclareAcronym{CMB}{
	short = CMB ,
	long = cosmic microwave background ,
	short-plural =  
}
\DeclareAcronym{DM}{
	short = DM ,
	long = dark matter ,
	short-plural =  
}

\DeclareAcronym{BBN}{
	short = BBN ,
	long = Big-Bang nucleosynthesis ,
	short-plural =  
}

\DeclareAcronym{SGWB}{
	short = SGWB ,
	long = stochastic gravitational	wave background ,
	short-plural =  s
}

\DeclareAcronym{LSS}{
	short = LSS ,
	long = large scale structure ,
	short-plural =  
}

\DeclareAcronym{RD}{
	short = RD ,
	long = radiation-dominated ,
	short-plural =  
}

\DeclareAcronym{BAO}{
	short = BAO ,
	long = baryon acoustic oscillations ,
	short-plural = 
}

\DeclareAcronym{ACT}{
	short = ACT ,
	long = Atacama Cosmology Telescope ,
	short-plural = 
}

\DeclareAcronym{SPT}{
	short = SPT ,
	long = South Pole Telescope  ,
	short-plural = 
}

\begin{document}

\title{Cosmological constraints on single-field inflation models from ACT/SPT data and observations at sub-Mpc scales}

\author{Qing-Yang Wang\orcidlink{0000-0001-9069-8737}} 
\affiliation{School of Physics, Sun Yat-Sen University, Guangzhou 510275, P. R. China}

\author{Jing-Zhi Zhou\orcidlink{0000-0003-2792-3182}} 
\email{zhoujingzhi@hau.edu.cn}
\affiliation{Department of Mathematics and Physics, Huai'an University, Meicheng East Road 1, Huai'an, Jiangsu 223200, China}

\author{Fei-Yu Chen\orcidlink{0000-0002-8674-9316}}
\affiliation{College of Physics and Materials Science, Tianjin Normal University, Tianjin 300387, China}

\author{Zhi-Chao Li\orcidlink{0009-0005-7984-2626}}
\affiliation{Center for Joint Quantum Studies and Department of Physics,
School of Science, Tianjin University, Tianjin 300350, China}

\begin{abstract}
We investigate single-field inflation models that can produce large curvature perturbations at sub-Mpc scales while remaining consistent with cosmological observations across multiple regimes. We highlight the tension between generating enhanced small-scale perturbations and maintaining agreement with the scalar spectral index $n_s$ measured by \ac{ACT} and \ac{SPT}, which disfavors several conventional single-field inflation models. To address this issue, we propose three viable classes of inflationary potentials: bump/dip-modified, step-like, and double-inflection-point. Each can sufficiently amplify primordial perturbations while predicting $n_s$ that lie within the $2\sigma$ range of current observational bounds. In addition, we examine the primordial power spectra predicted by these models and constrain their parameter spaces using cosmological observations at different scales, including \ac{CMB}, \ac{BAO}, \acp{PBH}, \ac{PTA}, and \ac{CMB}-distortion. Our results show that some of the models considered can simultaneously satisfy the ACT/SPT constraints and account for the current PTA observations, although doing so while respecting the constraints on the PBH abundance remains challenging.
\end{abstract}

\maketitle

\section{Introduction}
The inflation paradigm stands as a cornerstone of modern cosmology, successfully resolving longstanding puzzles such as the horizon, flatness, and monopole problems of the standard Big Bang model, while providing a compelling mechanism for generating the primordial perturbations that seed all subsequent cosmic evolution. Among its theoretical frameworks, single-field inflation has emerged as the most parsimonious and observationally favored scenario \cite{Baumann:2022mni,Guth:1980zm,Linde:1981mu,Germani:2017bcs,Di:2017ndc,Frolovsky:2025iao}. For a given single-field inflation model, one can compute the corresponding power spectrum of primordial curvature perturbations, which serves as the initial condition for subsequent cosmic evolution. 

By employing current cosmological observations, the physical properties of primordial curvature perturbations across different scales can be determined, thereby imposing stringent constraints on the parameter space of single-field inflation models. More precisely,  on large scales ($\gtrsim$1 Mpc), the amplitude of primordial curvature perturbations $A_{\zeta}$ is tightly constrained by \ac{CMB} and \ac{LSS} observations to around $A_{\zeta}\sim 2\times 10^{-9}$ \cite{Planck:2018vyg,Planck:2015fie,Planck:2015sxf}. Unlike the relatively precise cosmological observations on large scales, our understanding of the behavior of the primordial curvature perturbation power spectrum on small scales ($\lesssim$1 Mpc) remains limited \cite{Bringmann:2025cht}. When the primordial curvature perturbations are sufficiently large on small scales, they lead to the formation of \acp{PBH}, which are inevitably accompanied by the generation of \acp{SIGW} \cite{Saito:2008jc,Chang:2022nzu,Carr:2020xqk,Vaskonen:2020lbd,Saito:2009jt,Bugaev:2010bb,Baumann:2007zm,Alabidi:2012ex,Garcia-Bellido:2017aan,Cai:2018dig,Inomata:2018epa,Domenech:2019quo,Zhou:2020kkf,Adshead:2021hnm,Domenech:2021ztg,Vagnozzi:2023lwo,Firouzjahi:2023lzg,HosseiniMansoori:2023mqh,Mishra:2019pzq,Das:2023nmm,Cheung:2023ihl,Wang:2025kbj,Frosina:2023nxu,LISACosmologyWorkingGroup:2023njw,Mu:2023wdt,Fu:2022ypp,Kuang:2023urj,Terada:2025cto,Pi:2017gih,Wang:2024vfv,Wang:2025lti,Dvali:2025ktz,Balaji:2024hpu,Domenech:2024wao,Khlopov:2008qy,Wang:2025hwc,Kawai:2022emp,Lopez:2025gfu,Kawai:2023nqs,Chowdhury:2026vcf}. To better constrain the properties of primordial perturbations across different scales, it is necessary to combine cosmological data from multiple regimes. 

In June 2023, the \ac{PTA} collaborations NANOGrav \cite{NANOGrav:2023gor}, EPTA \cite{EPTA:2023fyk}, Parkes PTA \cite{Reardon:2023gzh}, and the China PTA \cite{Xu:2023wog} reported positive evidence for an isotropic, stochastic background of gravitational waves within the nHz frequency range, providing a crucial window into the evolution of the very early universe and into probing the physical properties of primordial perturbations at small scales. The origin of the observed nanohertz SGWB remains an open question. Potential sources include inspiraling supermassive black hole binaries (SMBHB) \cite{NANOGrav:2023hfp,Ellis:2023dgf,Shen:2023pan,Broadhurst:2023tus,Bi:2023tib,Kume:2024adn}, as well as various cosmological sources such as cosmic string \cite{Ellis:2023tsl,Kitajima:2023vre,Wang:2023len,Eichhorn:2023gat,Lazarides:2023ksx,Servant:2023mwt,Antusch:2023zjk,Yamada:2023thl,Ge:2023rce}, cosmological phase transition \cite{Zu:2023olm,Han:2023olf,Megias:2023kiy,Fujikura:2023lkn,Bai:2023cqj,Addazi:2023jvg,Xiao:2023dbb,Li:2023bxy,Ghosh:2023aum,DiBari:2023upq,Gouttenoire:2023bqy,Ahmadvand:2023lpp,An:2023jxf}, and \acp{SIGW} \cite{Franciolini:2023pbf,Cai:2023dls,Inomata:2023zup,Wang:2023ost,Liu:2023ymk,Ebadi:2023xhq,Abe:2023yrw,Yi:2023mbm,You:2023rmn,Balaji:2023ehk,Jin:2023wri,Zhu:2023gmx,Liu:2023pau,Jiang:2023gfe,Yi:2023tdk,Yuan:2023ofl,Chang:2023aba,Chang:2023vjk,Zhou:2024doz}. There are some comprehensive analyses of these sources \cite{NANOGrav:2023hvm,EPTA:2023xxk,Ellis:2023oxs,Figueroa:2023zhu,Bian:2025ifp}. The results indicate that \acp{SIGW}, originating from large primordial perturbations at small scales, possess the largest Bayesian factor. Therefore, compared with other possible sources, \acp{SIGW} are more likely to dominate the current \ac{PTA} observations. Assuming that \acp{SIGW} dominate the \ac{PTA} signals, the present \ac{PTA} data can be used to determine the energy density spectrum of \acp{SIGW}, thereby constraining the small-scale primordial power spectrum and the parameter space of the corresponding inflation models. Furthermore, even without assuming any specific origin of the \ac{SGWB}, the energy density spectrum of \acp{SIGW} cannot exceed the current observational upper limits, which also allows us to place constraints on specific inflation models.


In this study, we focus on single-field inflation models that may generate large primordial curvature perturbations at small scales, as well as the constraints imposed on these models by cosmological observations across different scales. Specifically, inflation model is typically driven by an inflaton field slow-rolling on an inflation potential. However, in certain inflation models, the slow-roll conditions may be temporarily violated due to subtle structures in the potential. This can cause a sharp amplification of curvature perturbations on certain scales, which in turn may serve as a source of \acp{SIGW}. To better constrain a single-field inflation model across different scales, we focus on the following cosmological observations: 

\noindent
(1) Constraints on the primordial power spectrum at $k\approx 10^{6}\mathrm{Mpc}^{-1}\sim10^{8}\mathrm{Mpc}^{-1}$ from PTA data under the assumption that \acp{SIGW} dominate the current \ac{PTA} observations \cite{NANOGrav:2023hvm}.
 
 \noindent
 (2) Constraints on the primordial power spectrum at $k\approx 1\mathrm{Mpc}^{-1}\sim10^{5}\mathrm{Mpc}^{-1}$
 from the CMB spectral distortion analysis of the FIRAS collaboration \cite{Wang:2025hbw,Fixsen:1996nj,Chluba:2012gq,Chluba:2012we,Chluba:2013dna,Bianchini:2022dqh,Yang:2024snb}.
 
 \noindent
 (3) Constraints on the total energy density of gravitational waves from \ac{CMB} and \ac{BAO} observations \cite{Wang:2025qpj,Cang:2022jyc,Ben-Dayan:2019gll,Zhou:2024yke}.
 
 \noindent
 (4) Constraints on the parameter space of inflation models from the abundance of \acp{PBH} \cite{DeLuca:2020agl,Madge:2023dxc,Karam:2022nym,Kawai:2021edk,Iovino:2024tyg,Yi:2023npi,Bhaumik:2022zdd}.

 It should be noted that when considering specific single-field inflation models, the enhancement of the primordial power spectrum at small scales may be constrained by the spectral index of the primordial power spectrum at large scales. Specifically, the ACT collaboration reports a scalar spectral index of $n_s=0.9743\pm0.0034$ at 68\% confidence levels based on a combination of Planck, ACT, BICEP/$Keck$, and DESI. A further analysis incorporating SPT data gives $n_s=0.9728\pm0.0027$ \cite{AtacamaCosmologyTelescope:2025blo,SPT-3G:2025bzu,Hai:2025wvs,Martin:2024qnn,Sabogal:2026qvy}. Many single-field inflation models with sufficiently amplified curvature perturbations at the PTA scale predict very lower $n_s$ values than these observational constraints. This tension arises because $n_s$ is generally positively correlated with the $e$-folding number during the slow-roll phase of inflation, and yet generating large curvature perturbations typically requires an ultra-slow-roll stage, which consumes considerable $e$-folds. As a result, the $e$-folds available for the conventional slow-roll phase are reduced, lowering the predicted value of $n_s$. For instance, the simplest inflection-point models typically predict $n_s=0.95-0.96$, which is obviously unfavorable with current observational bounds \cite{Garcia-Bellido:2017mdw,Ballesteros:2017fsr,Ezquiaga:2017fvi}. In this paper, we find three different types of single-field inflation models capable of generating large curvature perturbations at PTA scale while remaining consistent with the $n_s$ constraints from ACT and SPT: (1) bump/dip-modified potential, (2) step-like potential, and (3) double-inflection-point potential. All three classes can predict an appropriate $n_s$ that lie within the $2\sigma$ range of current observational bounds. We will scan the parameter space of these models, compute the resulting curvature perturbations and \acp{SIGW}, and constrain the parameter space of the aforementioned single-field inflation models using cosmological observations across different scales.

This paper is organized as follows. In Sec.~\ref{sec:models} we introduce three classes of single-field inflation models that can enhance the primordial power spectrum at small scales while remaining consistent with the \ac{ACT}/\ac{SPT} constraint on $n_s$, and we compute the corresponding primordial power spectra numerically. In Sec.~\ref{sec:obs} we summarize the cosmological observables used in this work: the energy-density spectrum of \acp{SIGW}, the \ac{CMB} spectral distortions, and the abundance of \acp{PBH}. In Sec.~\ref{sec:results} we scan the parameter space of each model and map out the regions allowed by these observations. In Sec.~\ref{sec:PTA} we perform a Bayesian analysis of the NANOGrav 15-year data set to examine whether the \acp{SIGW} produced by these models can dominate the current \ac{PTA} signal within the cosmologically allowed parameter space. Our conclusions and discussions are presented in Sec.~\ref{sec:conclusion}.

\section{Single-field inflation models}\label{sec:models}
In this section, we investigate several different single-field inflation models and calculate the corresponding power spectra of primordial curvature perturbations. We begin with a minimally coupled single-field inflation action
\begin{equation}
    S=\int\mathrm{d}^4x\sqrt{-g}\left[\frac{M_P^2}{2}R-\frac{1}{2}\partial_\mu\phi\partial^\mu\phi-V(\phi)\right] \ .
\end{equation}
The inflaton field is generally regarded as an average background with a slight fluctuation, $\phi(t,\mathbf x)\rightarrow\phi(t)+\delta\phi(t,\mathbf x)$. The line element of perturbed \ac{FLRW} spacetime can be expressed as
\begin{eqnarray}
    \mathrm{d} s^2&=&-\left(1+2A\right)\mathrm{d} t^2+2a(t)\partial_iB\mathrm{d} x^i\mathrm{d} t \nonumber\\
    &+& a^2(t)\delta_{ij}\mathrm{d} x^i\mathrm{d} x^j \ ,
\end{eqnarray}
where $A(t,\mathbf x)$ and $B(t,\mathbf x)$ describe the perturbations related to $\delta\phi$. Varying the action with this metric, the equations of motion for the inflaton background and perturbations can be derived as
\begin{eqnarray}
    &&\ddot\phi+3H\dot\phi+V'(\phi)=0 \ ,\label{EOM}\\
    &&\ddot{\delta\phi}+3H\dot{\delta\phi}-\frac{\nabla^2}{a^2}\delta\phi=\left[\left(3+2\epsilon-\eta\right) \right. \nonumber\\
    &&~~~~~~~~~~~~~~~~~~\left.\times \left(\epsilon-\eta\right)-\frac{\dot\eta}{H}\right]H^2\delta\phi  \ ,\label{EOMper}
\end{eqnarray}
where $H\equiv\dot a/a$ is the Hubble parameter, and we have introduced the slow-roll parameters
\begin{gather}
    \epsilon\equiv-\frac{\dot H}{H^2} \ ,~~~\eta\equiv-\frac{\ddot\phi}{H\dot\phi} \ .
\end{gather}
The curvature perturbation $\zeta$ has the following relation with the inflaton perturbation
\begin{align}
    \zeta=-\frac{H}{\dot\phi}\delta\phi \ .
\end{align}
Substituting it into Eq.~(\ref{EOMper}) and transforming the equation into Fourier space, we derive the equation of motion of  curvature perturbation
\begin{equation}
    \ddot{\zeta_k}+\left(3+2\epsilon-2\eta\right)H\dot \zeta_k+\frac{k^2}{a^2}\zeta_k=0  \ .
\end{equation}

For practicality, we convert the variable $\mathrm{d} t$ to the $e$-folding number, $\mathrm{d} N=H\mathrm{d} t$, then the above equations are rewritten as
\begin{gather}
    \phi_{,NN}+\left(3-\epsilon\right)\phi_{,N}+\frac{V_{,\phi}}{H^2}=0  \ ,\\
    \zeta_{k,NN}+\left(3+\epsilon-2\eta\right) \mathcal \zeta_{k,N}+\frac{k^2}{H^2e^{2N}}\zeta_k=0  \ ,\\
    \epsilon=\frac{\phi_{,N}^2}{2M_P^2} \ ,~~~~~\eta=\epsilon-\frac{\phi_{,NN}}{\phi_{,N}}  \ .
\end{gather}
Combining these equations with the Friedmann equation for inflaton dominated phase
\begin{align}
    H^2=\frac{V}{\left(3-\epsilon\right)M_P^2} \ ,
\end{align}
we can solve the equation of motion of primordial curvature perturbation, and derive the corresponding primordial power spectrum for super-horizon modes
\begin{align}
    \mathcal{P}_\zeta(k)=\frac{k^3}{2\pi^2}|\zeta_k|^2\ .
\end{align}

The following parts of this chapter are devoted to applying the aforementioned results to concrete single-field inflationary models. As we mentioned earlier, we find three competitive single-field inflation potentials: (1) bump/dip-modified potential, (2) step-like potential, and (3) double-inflection-point potential. They are capable of amplifying curvature perturbations to $\mathcal{P}_\zeta\approx10^{-2}$ at the PTA scale ($k\approx 10^7~\mathrm{Mpc}^{-1}$) while remaining consistent with the $n_s$ constraints at the CMB scale. 

\subsection{Bump/dip-modified potential}
The bump/dip-modified potential can be written as
\begin{align}
    V(\phi)=V_b(\phi)\left[1+V_\delta(\phi)\right] \ ,
\end{align}
where $V_b(\phi)$ refers to a background potential, and $\delta(\phi)$ is a little bump/dip modification on it. Following previous studies \cite{Mishra:2019pzq,Das:2023nmm,Cheung:2023ihl,Wang:2025kbj}, we set it in a Gaussian form 
\begin{align}\label{bump}
    V_\delta(\phi)=A\exp\left[-\frac{(\phi-\phi_0)^2}{2\sigma^2}\right] \ ,
\end{align}
where $A$, $\sigma$, and $\phi_0$ describe the height, width, and position of the Gaussian modification, respectively. If $A>0$, it is called a bump, else it is a dip. When these parameters are appropriate, the bump/dip can trap the inflation field for several e-folds during its evolution, thereby achieving the ultra-slow-roll stage and amplify curvature perturbations. The background potential can be arbitrary, but in order to comply with the ACT/SPT observational constraints, it is best for the background potential to provide a spectral index of $n_s\gtrsim0.975$ for $N=60$ without any bump/dip modifications. The reason is as we mentioned earlier: the ultra-slow-roll phase generated by the bump/dip term will occupy several e-folds and shorten the slow-roll phase, thus making the predicted $n_s$ smaller than that without the bump/dip. Moreover, the predicted tensor-to-scalar ratio $r$ is also required to be lower than the current constraint from the BICEP/$Keck$ collaboration: $r<0.036$ at 95\% confidence levels.

One of the widely discussed models that meets these conditions is the Kachru-Kallosh-Linde-Trivedi (KKLT) model, which is inspired by string theory. Its potential can be written as
\begin{align}\label{VKKLT}
    V_b(\phi)=V_0\frac{1}{1+|M/\phi|^n} \ ,
\end{align}
where $M$ and $n$ are free parameters, and $V_0$ is fixed by the intensity of curvature perturbations observed on the CMB pivot scale. Assuming $M\lesssim1$, the spectral index $n_s$ and the tensor-to-scalar $r$ of primordial perturbations at the CMB pivot scale ($N\sim[50,60]$ before inflation ends) can be approximately derived
\begin{align}
    n=1:~&n_s\simeq1-\frac{4}{3N}-\frac{M^{2/3}}{(3N)^{4/3}},~r\simeq\frac{8M^{2/3}}{(3N)^{4/3}} \ ,\\
    n=2:~&n_s\simeq1-\frac{3}{2N}-\frac{M}{(2N)^{3/2}},~r\simeq\frac{\sqrt2M}{N^{3/2}} \ ,\\
    n\geq3:~&n_s\simeq1-\frac{2(n+1)}{(n+2)N_\mathrm{cor}}\ ,
\end{align}
\begin{equation}
r\approx\frac{8n}{(n+2)N_\mathrm{cor}}\left(\frac{M^2}{n(n+2)N_\mathrm{cor}}\right)^{\frac{n}{n+2}} \ ,
\end{equation}
where $N_\mathrm{cor}=N-\frac{1}{2n}\left[n(n+2)NM^n\right]^{\frac{2}{n+2}}$ is a corrected $e$-folding number. It is seen that $n_s\sim0.97-0.977$ for a wide range of parameters, which aligns well with the latest observations. In this paper, we set $n=1$ and $M=1$.

\begin{figure*}[!t]
    \centering\includegraphics[width=1\textwidth]{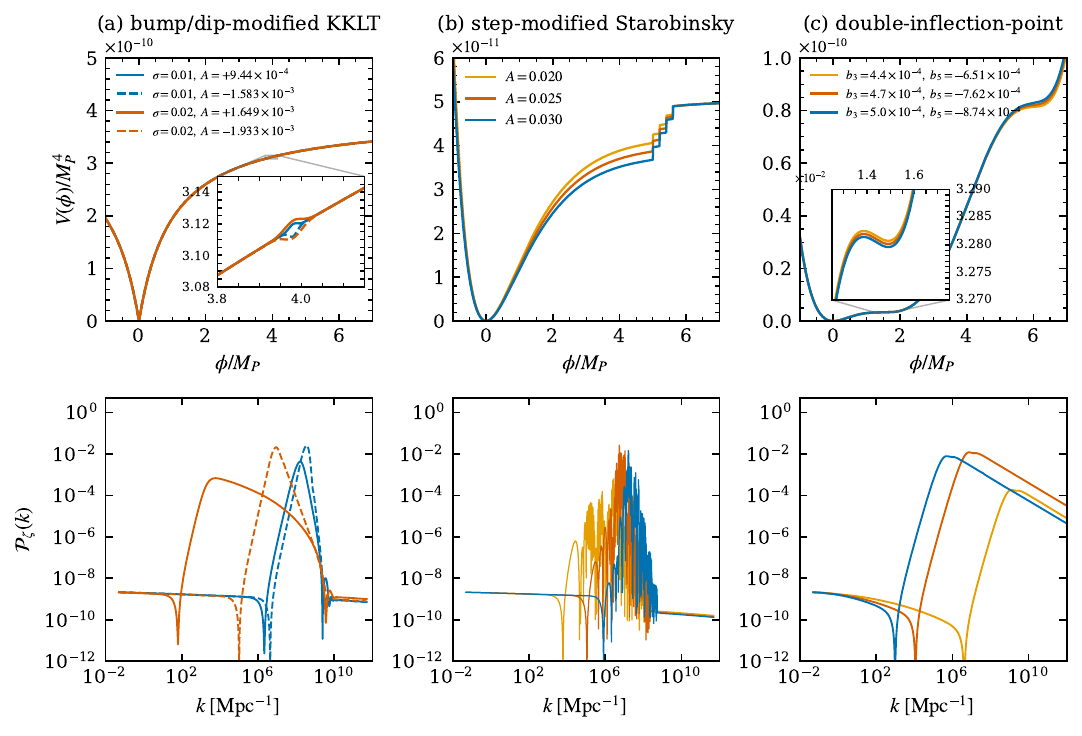}
    \caption{Inflation potentials (upper panels) and curvature perturbation spectra $\mathcal{P}_\zeta(k)$ (lower panels) for three inflation models. Left: the bump/dip-modified KKLT model with $M=1$, $n=1$, $V_0=3.9\times10^{-10}M_P^4$, $\phi_0=3.98M_P$ fixed; solid (dashed) lines correspond to positive (negative) feature amplitude $A$, and the inset zooms into the bump/dip region of the potential. Middle: the step-modified Starobinsky model with $V_0=5\times10^{-11}M_P^4$, $n=3$, $\sigma=10^3$, $\phi_0=5$, $d=0.2$ fixed; the rapid oscillations of the spectra around the peak are physical and are shown without any smoothing. Right: the double-inflection-point model with $\Lambda=1$, $V_0=5.7\times10^{-12}M_P^4$, $\phi_1=1.437$, $\phi_2=6$, $b_2=0$, $b_6=-10^{-3}$ fixed; the inset zooms into the ultra-slow-roll dip of the potential.\label{fig1}}
\end{figure*}

After specifying the inflation model, we calculate the curvature perturbation spectrum generated by the background potential given in Eq.~(\ref{VKKLT}) with a Gaussian bump/dip term in Eq.~(\ref{bump}). The left panel of Fig.~\ref{fig1} displays the potential function curve for the considered inflation model along with its primordial power spectrum. More precisely, the left-upper panel shows the inflation potential, where it is clear that the bump/dip term has a negligible influence on the configuration of the background potential. Nevertheless, it induces a sufficiently intense peak in the curvature perturbation spectrum that can generate sizable GWs. The position and intensity of the peak are mainly controlled by the parameters $\phi_0$ and $A$, respectively, while the parameter $\sigma$ affects both the intensity and the width of the peak. In the subsequent sections, we will fix $\sigma$ and investigate the impact of the remaining two parameters on the resulting GW spectrum.

\subsection{Step-like potential}
The step-like potential also involves modifications on a background potential
\begin{align}
    V(\phi)=V_b(\phi)\left[1+V_s(\phi)\right],
\end{align}
where the step term is considered as follows \cite{Dalianis:2021iig,Mastache:2023cha,Stamou:2024lqf}
\begin{align}
    V_s(\phi)=A\sum^n_{m=0}\left(-1+\tanh{\left[\sigma\left(\phi-\phi_0-md\right)\right]}\right).
\end{align}
The properties of the steps, namely, their number, position, height, and width are controlled by the parameters $n$, $\phi_0$, $A$, and $d$, respectively, while the parameter $\sigma$ regulates the smoothness of the step edges. Such step terms can naturally arise in the $\alpha$-attractor model in supergravity. Compared to other single-field inflation mechanisms that enhance curvature perturbations, the step mechanism notably alleviates the fine-tuning problem, offering a compelling theoretical motivation. The presence of the step terms actually leads to a larger variation of the inflaton field $\phi$ over the same $e$-folds, which generally results in an increase in the spectral index $n_s$. Consequently, some inflation models previously disfavored by ACT/SPT observations due to overly low $n_s$ predictions may become consistent with current observational constraints within the $2\sigma$ range once the step term is incorporated. We consider the Starobinsky potential as the background potential
\begin{align}
    V_b(\phi)=V_0\left(1-e^{-\sqrt{2/3}\phi}\right)^2.
\end{align}
The spectral index $n_s$ and the tensor-to-scalar $r$ of this potential are approximately derived as follows 
\begin{align}
    n_s\simeq1-\frac{2}{N+3}-\frac{3}{(N+3)^2},~r\simeq\frac{12}{(N+3)^2}.
\end{align}
For $N\sim[50,60]$, it predicts $n_s=0.961-0.967$, slightly below the $2\sigma$ lower limit of observational constraints. However, after introducing the step terms, it gives a result consistent with the observations. For example, setting parameters $V_0=5\times10^{-11}M_P^4$, $n=3$, $A=0.025$, $\sigma=10^3$, $\phi_0=5$, $d=0.2$, and the $e$-folding number $N=55$, the spectral index is predicted as $n_s=0.977$.

The middle panel in Fig.~\ref{fig1} illustrates examples of the step-like potential and the corresponding curvature perturbation spectrum. We introduce four steps to raise the peak of the spectrum to $\mathcal {P}_\zeta(k)\sim10^{-2}$. While the scale and smoothness of the steps, controlled by $A$, $d$, and $\sigma$, also influence the peak intensity of $\mathcal{P}_\zeta(k)$, they are difficult to further raise the upper limit of $\mathcal {P}_\zeta(k)$ in this four-step configuration.

\subsection{Double-inflection-point potential}
The ordinary single-inflection-point inflation model is a widely studied scenario capable of enhancing curvature perturbations \cite{Garcia-Bellido:2017mdw,Ballesteros:2017fsr,Ezquiaga:2017fvi,Ballesteros:2020qam,Barker:2024mpz}. However, it typically predicts a spectral index $n_s$ that falls below the observational constraint interval, since its potential tends to flatten in the large $\phi$ region. As an extension of this framework, the double-inflection-point model introduces a second inflection point to elevate the potential at large $\phi$ \cite{Xu:2019bdp,Gao:2021lno,Frosina:2023nxu,Allegrini:2024ooy}, thereby bringing $n_s$ into better agreement with observational bounds. Here, we consider a polynomial potential up to $n=6$
\begin{align}\label{Vdouble}
    V(\phi)=V_0\sum_{n=2}^6c_n\left(\frac{\phi}{\Lambda}\right)^n \ .
\end{align}
The necessary condition for an inflection point $\phi_i$ is $V''(\phi_i)=0$. Generally, to produce significant curvature perturbations, the field $\phi$ must evolve slowly near the inflection point, which requires $V'(\phi_i)\to0$. In the case of two inflection points $\phi_{1}$ and $\phi_2$, if we strictly require $V'(\phi_{1,2})=V''(\phi_{1,2})=0$ and fix $c_4=1$, the other coefficients in Eq.~(\ref{Vdouble}) can be solved as
\begin{gather}
    c_2=2\phi_1^2\phi_2^2\kappa,~c_3=-\frac{8}{3}\phi_1\phi_2\left(\phi_1+\phi_2\right)\kappa,\nonumber\\
    c_5=-\frac{8}{5}\left(\phi_1+\phi_2\right)\kappa,~c_6=\frac{2}{3}\kappa,
\end{gather}
where $\kappa\equiv1/\left(\phi_1^2+4\phi_1\phi_2+\phi_2^2\right)$. However, since the condition $V'(\phi_{1,2})=0$ is not strict in practice, these coefficients may slightly deviate from the expressions above. We account for this by introducing small corrections:  $c_n\to c_n\left(1+b_n\right)$. Accordingly, the potential in Eq.~(\ref{Vdouble}) can be reformulated as
\begin{align}
    V(\phi)=V_0\bigg[&\phi^4+\kappa\bigg(2\phi_1^2\phi_2^2\left(1+b_2\right)\phi^2\nonumber\\
    &-\frac{8}{3}\phi_1\phi_2\left(\phi_1+\phi_2\right)\left(1+b_3\right)\phi^3\nonumber\\
    &-\frac{8}{5}\left(\phi_1+\phi_2\right)\left(1+b_5\right)\phi^5 \\
    &+\frac{2}{3}\left(1+b_6\right)\phi^6\bigg)\bigg].
\end{align}
This form mitigates the fine-tuning problem and is more suitable for numerical analysis compared to the original Eq.~(\ref{Vdouble}).

We illustrate several examples of this model in the right panel of Fig.~\ref{fig1}. In comparison with the bump/dip-modified potentials, the ultra-slow-roll phase in this model lasts significantly longer, leading to a more gradual decrease in the curvature perturbation spectrum on small scales. Furthermore, the spectral index $n_s$ exhibits a strong dependence on the total $e$-folds $N$, which must therefore be carefully adjusted to bring $n_s$ into agreement with observational constraints. The parameter $\phi_1$ determines the position of the inflection point associated with the ultra-slow-roll stage, thereby controlling the peak position of the curvature perturbation spectrum. While other parameters also influence the spectrum. For the sake of alleviating fine-tuning problem, we choose to focus on $b_3$ and $b_5$ in subsequent sections to adjust both the position and amplitude of the spectral peak.

\section{Cosmological observations}\label{sec:obs}
In the preceding sections, we examined three distinct single-field inflation models and derived their corresponding primordial power spectra. In this section, we turn to the constraints imposed by cosmological observations at different scales on these spectra and on the parameter space of the associated inflation models.

\subsection{Scalar induced gravitational waves}
The investigation of \acp{SIGW} on sub-Mpc  scales is motivated by studies of second‑order tensor and vector perturbations generated by first‑order scalar modes and their impact on \ac{CMB} polarization \cite{Mollerach:2003nq}. Building on earlier analyses of second‑order perturbations at \ac{CMB} scales, Refs.~\cite{Ananda:2006af,Osano:2006ew,Baumann:2007zm} examined second‑order \acp{GW} induced by large‑amplitude primordial scalar perturbations on small scales and computed the associated energy density spectrum. Over the past decade, research on \acp{SIGW} has expanded into a wide range of directions \cite{Domenech:2021ztg}. Furthermore, for \acp{SIGW}, their generation is guaranteed whenever primordial perturbations are present, and they inevitably leave direct or indirect imprints on cosmological observations at the corresponding scales. Specifically, by deriving the second‑order cosmological perturbation equations on an \ac{FLRW} background and extracting their transverse–traceless component, one obtains \cite{Malik:2008im,Zhou:2024doz}
\begin{equation}\label{eq:h}
	\begin{aligned}
		h_{ij}^{(2)''}(\eta,\mathbf{x})&+2 \mathcal{H} h_{ij}^{(2)'}(\eta,\mathbf{x})-\Delta h_{ij}^{(2)}(\eta,\mathbf{x}) \\
        &=-4 \Lambda_{ij}^{lm} \mathcal{S}^{(2)}_{lm}(\eta,\mathbf{x})  \ ,
	\end{aligned}
\end{equation}
where $\Lambda_{ij}^{lm}$ is the decomposed operator to extract the transverse and traceless terms \cite{Chang:2020tji}. The source term in Eq.~(\ref{eq:h}) is given by
\begin{eqnarray}
	\mathcal{S}^{(2)}_{lm}(\eta,\mathbf{x})&=&\partial_{l} \phi^{(1)} \partial_{m} \phi^{(1)} +4 \phi^{(1)} \partial_{l} \partial_{m} \phi^{(1)}\nonumber\\
    &-&\frac{1}{ \mathcal{H}}\left(\partial_{l} \phi^{(1)'} \partial_{m} \phi^{(1)}+\partial_{l} \phi^{(1)}  \partial_{m} \phi^{(1)'}\right) \nonumber\\
	&-&\frac{1}{ \mathcal{H}^{2}} \partial_{l} \phi^{(1)'} \partial_{m}  \phi^{(1)'} \ .
	\label{eq:1} 
\end{eqnarray}
Eq.~(\ref{eq:h}) demonstrates that first‑order scalar perturbations couple to second‑order tensor modes within the second‑order cosmological perturbation equations. Therefore, any primordial scalar perturbation produced during inflation necessarily gives rise to second‑order \acp{SIGW}. In this paper, we concentrate on \acp{SIGW} in the nHz band and examine how \ac{PTA}, \ac{CMB}, and \ac{BAO} measurements constrain the parameter space of inflation models. As an important component of the \ac{SGWB}, the energy density spectrum of \acp{SIGW} is encoded in the following double‑integral form \cite{Espinosa:2018eve,Kohri:2018awv}
\begin{align}
    \Omega_{\mathrm{GW}}^{(2)}(k) &= \int_{0}^{\infty} \mathrm{d}v \int_{|1-v|}^{1+v} \mathrm{d}u\,
    \mathcal{P}_{\zeta}(uk) \mathcal{P}_{\zeta}(vk) \notag \\
    &\times  \frac{3}{1024 u^8 v^8} (u^2 + v^2 - 3)^2 \notag \\
    &\times \left[4v^2 - (1 + v^2 - u^2)^2\right]^2 \notag \\
    &\times \left\{ \left[(u^2 + v^2 - 3) \ln \left| \frac{3 - (u + v)^2}{3 - (u - v)^2} \right| - 4uv \right]^2 \right. \notag \\
    & + \left.  \pi^2 (u^2 + v^2 - 3)^2 \Theta \Big(u + v - \sqrt{3} \Big) \right\} \ ,
    \label{eq:1Oss}
\end{align}
where $\mathcal{P}_{\zeta}$ denotes the power spectrum of primordial curvature perturbation computed in the previous section. It is evident that the energy density spectrum of \acp{SIGW} scales proportionally to the square of the amplitude of primordial power spectrum. Consequently, only sufficiently large primordial curvature perturbations can generate \acp{SIGW} with observable signatures. In addition, Eq.~(\ref{eq:1Oss}) describes the energy density spectrum during the \ac{RD} era. Taking into account the thermal history of the universe, the current total energy density spectrum $\bar{\Omega}^{(2)}_{\mathrm{GW,0}}(k)$ is given by
\begin{equation}
\bar{\Omega}^{(2)}_{\mathrm{GW,0}}(k) = \Omega_{\mathrm{rad},0}\left(\frac{g_{*,\rho,e}}{g_{*,\rho,0}}\right)\left(\frac{g_{*,s,0}}{g_{*,s,e}}\right)^{4/3}\bar{\Omega}^{(2)}_{\mathrm{GW}}(k) \ ,
\end{equation}
where $\Omega_{\mathrm{rad},0}$ ($ =4.2\times 10^{-5}h^{-2}$) is the energy density fraction of radiation today.

In addition to the direct measurement of the energy‑density spectrum of \acp{SIGW}, the \acp{SIGW} contribute as an additional radiation component and thus affect large‑scale cosmological observations \cite{Wang:2025qpj,Cang:2022jyc,Ben-Dayan:2019gll,Zhou:2024yke}. Specifically, the total energy density of \acp{SIGW} must satisfy
\begin{equation}\label{eq:rhup}
h^2\rho_{\mathrm{GW}}=\int_{f_{\mathrm{min}}}^{\infty} h^2\Omega_{\mathrm{GW},0}(k) \mathrm{d}\left(\ln k\right) < 2.9\times 10^{-7}  \ ,
\end{equation}
at $95\%$ confidence level for \ac{CMB}$+$\ac{BAO} data \cite{Clarke:2020bil}. 

\begin{figure*}[!t]
    \centering\includegraphics[width=1\textwidth]{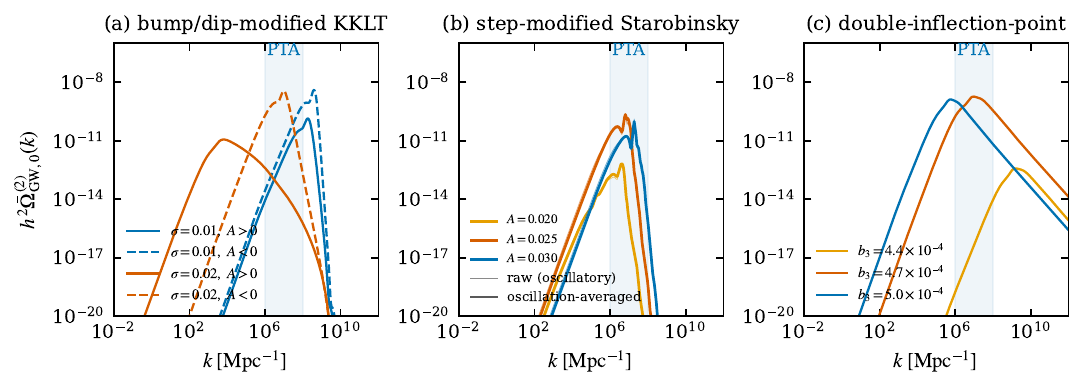}
    \caption{Present-day energy-density spectra $h^2\bar{\Omega}_{\mathrm{GW},0}^{(2)}(k)$ of the \acp{SIGW} produced by the enhanced primordial spectra of the three inflation models, for the same parameter sets as in Fig.~\ref{fig1}. Left: bump/dip-modified KKLT model. Middle: step-modified Starobinsky model; the thin light curves are computed from the raw oscillatory primordial spectra, while the thick curves use the oscillation-averaged spectra (arithmetic average over a $0.1$~dex window in $\ln k$). Right: double-inflection-point model. The shaded vertical band marks the \ac{PTA}-sensitive range.\label{fig:sigw}}
\end{figure*}

We numerically evaluate the double integral in Eq.~(\ref{eq:1Oss}), together with the thermal-history correction, for the primordial power spectra obtained in the previous section. Figure~\ref{fig:sigw} shows the resulting present-day energy-density spectra $h^2\bar{\Omega}_{\mathrm{GW},0}^{(2)}(k)$ for the benchmark parameter sets of Fig.~\ref{fig1}. The spectra peak in or close to the \ac{PTA} band, with peak values $h^2\bar{\Omega}_{\mathrm{GW},0}^{(2)}\simeq3.6\times10^{-9}$ at $k\simeq10^{7}\,\mathrm{Mpc}^{-1}$ for the bump/dip-modified KKLT model ($\sigma=0.02$, $A<0$), $\simeq2.3\times10^{-10}$ at $k\simeq6\times10^{6}\,\mathrm{Mpc}^{-1}$ for the step-modified Starobinsky model ($A=0.025$), and $\simeq1.85\times10^{-9}$ at $k\simeq10^{7}\,\mathrm{Mpc}^{-1}$ for the double-inflection-point model ($b_3=4.7\times10^{-4}$). The spectra of the step-modified model inherit the rapid oscillations of its primordial spectrum; in Fig.~\ref{fig:sigw}(b) we therefore display both the spectra computed from the raw primordial spectra (thin light curves) and those from oscillation-averaged primordial spectra (thick curves, arithmetic average over a $0.1$~dex window in $\ln k$). 

\subsection{CMB-distortion}
\ac{CMB} spectral distortions provide an independent and complementary probe of the primordial curvature perturbation on scales smaller than those directly constrained by CMB observations. Small-scale curvature perturbations dissipate their energy through photon diffusion after entering the horizon, leading to energy injection into the primordial plasma and consequently generating spectral distortions of the CMB blackbody spectrum.
At redshifts $z\lesssim 10^6$, energy injection into the primordial plasma can leave persistent spectral distortions in the CMB. At early times, the resulting distortion is predominantly of the chemical potential $\mu$-type, while at lower redshifts, $z\lesssim5\times10^4$, Compton $y$-type distortions are generated. In this work, we focus on the conventional $\mu$-type and $y$-type spectral distortions.
For a given primordial curvature power spectrum $P_\zeta(k)$, the spectral distortions can be expressed as \begin{equation} X=\int_{k_{\min}}^\infty\frac{dk}{k}  P_\zeta(k)W_X(k) \ , \qquad X=\mu,y \ , \label{eq:CMB_distortion_general} \end{equation} where $k_{\min}=1,{\rm Mpc}^{-1}$ and $W_X(k)$ denotes the corresponding window function. Following Refs.~\cite{Wang:2025hbw,Fixsen:1996nj,Chluba:2012gq,Chluba:2012we,Chluba:2013dna,Bianchini:2022dqh,Yang:2024snb,Tagliazucchi:2023dai}, the window functions can be approximated as \begin{equation}\label{eq:W_mu} 
\begin{aligned}
W_\mu(k)&= \frac{2.2\exp\left[-(\hat{k}/1360)^2\right]} {1+(\hat{k}/260)^{0.6}+\hat{k}/340} \\
&~~~~~~~~~~~-2.2\exp\left[-(\hat{k}/32)^2\right] \ , 
\end{aligned}
\end{equation} 
and 
\begin{equation} W_y(k)=0.4\exp\left[-(\hat{k}/32)^2\right]\ , \label{eq:W_y} \end{equation} 
where $\hat{k}\equiv k/1{\rm Mpc}^{-1}$. The COBE/FIRAS observations constrain the spectral distortions at $95\%$ confidence level as \begin{equation} \mu\leq4.7\times10^{-5}, \qquad y\leq1.5\times10^{-5}. \label{eq:FIRAS} \end{equation} These bounds can therefore be used to constrain the primordial curvature power spectrum over the range of scales to which the corresponding window functions are sensitive.

\subsection{Primordial black holes}
Small‑scale primordial curvature perturbations with sufficiently large amplitudes induce pronounced density contrasts that seed the formation of \acp{PBH}. Observational upper limits on \ac{PBH} abundance therefore constrain the properties of the primordial curvature power spectrum on small scales. In this subsection we apply threshold theory to compute \ac{PBH} production in various single‑field inflation models and use current \ac{PBH} observational bounds to restrict the primordial power spectrum on sub‑Mpc scales and the corresponding parameter space of inflation models.

In threshold statistics, the \ac{PBH} abundance is computed by integrating the probability distribution function of the smoothed density contrast from the threshold value $\mathcal{C}_{\mathrm{th}}$, defined as
\begin{eqnarray}\label{eq:betat}
    \beta_{\mathrm{th}}=\int_{\mathcal{D}} \mathcal{K}\left(\mathcal{C}-\mathcal{C}_{\mathrm{th}}\right)^\gamma \mathrm{P}_{\mathrm{G}}\left(\mathcal{C}_{\mathrm{G}}, \zeta_{\mathrm{G}}\right) \mathrm{d} \mathcal{C}_{\mathrm{G}} \mathrm{~d} \zeta_{\mathrm{G}} \ ,
\end{eqnarray}
where the domain of integration is $\mathcal{D}=$ $\left\{\mathcal{C}\left(\mathcal{C}_{\mathrm{G}}, \zeta_{\mathrm{G}}\right)>\mathcal{C}_{\text {th}} \wedge \mathcal{C}_1\left(\mathcal{C}_{\mathrm{G}}, \zeta_{\mathrm{G}}\right)<2 \Phi\right\}$. In this paper, we set $\mathcal{K}=4.4$ and $\gamma=0.38$ \cite{Musco:2023dak,Iovino:2024tyg}. Furthermore, the compaction function $\mathcal{C}=\mathcal{C}_1-\mathcal{C}_1^2 /(4 \Phi)$ can be obtained from the linear $\mathcal{C}_1=\mathcal{C}_{\mathrm{G}} \mathrm{d} F / \mathrm{d} \zeta_{\mathrm{G}}$ component, that uses $\mathcal{C}_{\mathrm{G}}=$ $-2 \Phi r \zeta_G^{\prime}$, where $\Phi=3(1+w)/(5+3w)$ and $\zeta=F(\zeta_{G})$. When considering Gaussian primordial curvature perturbations, we have $F(\zeta_G)=\zeta_G$. The Gaussian components in Eq.~(\ref{eq:betat}) are distributed as
\begin{equation}\label{eq:PG1}
    P_{\mathrm{G}}\left(\mathcal{C}_{\mathrm{G}}, \zeta_{\mathrm{G}}\right)=\frac{e^{\left[-\frac{1}{2\left(1-\gamma_{c r}^2\right)}\left(\frac{\mathcal{C}_{\mathrm{G}}}{\sigma_c}-\frac{\gamma_{c r} \zeta_{\mathrm{G}}}{\sigma_r}\right)^2-\frac{\zeta_\mathrm{G}^2}{2 \sigma_r^2}\right]}}{2 \pi \sigma_c \sigma_r \sqrt{1-\gamma_{c r}^2}} \ ,
\end{equation}
where the correlators in Eq.~(\ref{eq:PG1}) are given by
\begin{equation}\label{eq:sgm3}
\begin{aligned}
& \sigma_c^2=\frac{4 \Phi^2}{9} \int_0^{\infty} \frac{\mathrm{d} k}{k}\left(k r_m\right)^4 W^2\left(k, r_m\right) P_\zeta^T \ , \\
& \sigma_{c r}^2=\frac{2 \Phi}{3} \int_0^{\infty} \frac{\mathrm{d} k}{k}\left(k r_m\right)^2 W\left(k, r_m\right) W_s\left(k, r_m\right) P_\zeta^T \ , \\
& \sigma_r^2=\int_0^{\infty} \frac{\mathrm{d} k}{k} W_s^2\left(k, r_m\right) P_\zeta^T \ ,
\end{aligned}
\end{equation}
where $P_\zeta^T=T^2\left(k, r_m\right) P_\zeta(k)$, and $\gamma_{c r} \equiv \sigma_{c r}^2 / \sigma_c \sigma_r$. Here, we have defined $W\left(k, r_m\right), W_s\left(k, r_m\right)$ and $T\left(k, r_m\right)$ as the top-hat window function, the spherical-shell window function, and the radiation transfer function \cite{Young:2022phe}. Using Eq.~(\ref{eq:betat}) together with a specified primordial power spectrum, the resulting \ac{PBH} abundance can be written as \cite{Sasaki:2018dmp}
\begin{equation}\label{eq:pbh1}
\begin{aligned}
 f_{\mathrm{pbh}} \simeq 2.5 \times 10^8 \beta_{\mathrm{th}}\left(\frac{g_*^{\text {form }}}{10.75}\right)^{-\frac{1}{4}}\left(\frac{m_{\mathrm{pbh}}}{M_{\odot}}\right)^{-\frac{1}{2}} \ .
\end{aligned}
\end{equation}
In this paper we restrict ourselves to Gaussian primordial curvature perturbations, i.e. $F(\zeta_\mathrm{G})=\zeta_\mathrm{G}$. The threshold $\mathcal{C}_{\mathrm{th}}$ is evaluated with the shape-dependent prescription of Ref.~\cite{Musco:2023dak}, and the smoothing scale $r_m$ is chosen to maximize the compaction function. 

\section{Observational constraints on the inflation models}
\label{sec:results}

\begin{figure*}[!t]
    \centering\includegraphics[width=1\textwidth]{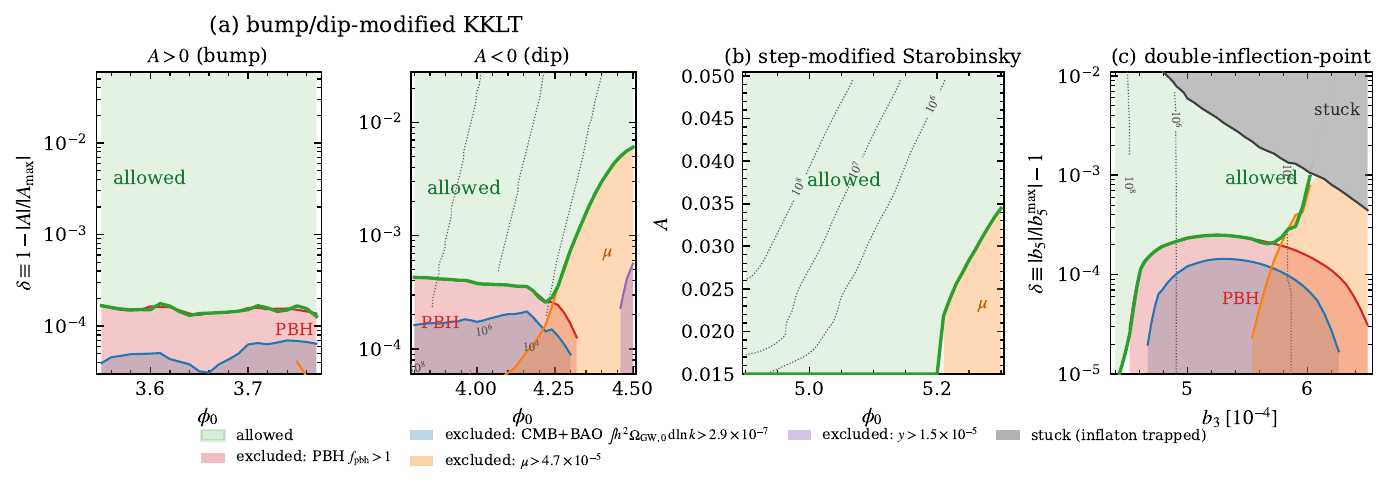}
    \caption{Observational constraints on the parameter spaces of the three inflation models. (a) Bump/dip-modified KKLT model with $\sigma=0.02$, shown in the $(\phi_0,\,\delta)$ plane with $\delta\equiv 1-|A|/|A_{\max}|$ for the bump ($A>0$, left) and dip ($A<0$, right) cases; (b) step-modified Starobinsky model with $n=3$, $\sigma=10^3$, $d=0.2$, shown in the $(\phi_0,\,A)$ plane; (c) double-inflection-point model, shown in the $(b_3,\,\delta)$ plane with $\delta\equiv|b_5|/|b_5^{\max}|-1$. In every panel the light-green region labeled ``allowed'' satisfies all the constraints considered in this work. The colored regions are excluded: red by the \ac{PBH} abundance ($f_{\mathrm{pbh}}>1$); blue by the \ac{CMB}$+$\ac{BAO} bound on the total \ac{SIGW} energy density, Eq.~\eqref{eq:rhup}; orange and purple by the COBE/FIRAS $\mu$- and $y$-distortion bounds, Eq.~\eqref{eq:FIRAS}, respectively. In panel (c) the dark-gray region is excluded because the inflaton is trapped in the ultra-slow-roll dip and cannot reach the potential minimum ($|b_5|>|b_5^{\max}|$). In panel (b) the \ac{PBH}, \ac{SIGW}, and $y$-distortion constraints are all weaker than the $\mu$-distortion bound and do not appear. The dotted gray contours mark the peak scale $k_p=10^{4},\,10^{6},\,10^{8}\,\mathrm{Mpc}^{-1}$ of the primordial power spectrum. All distortion constraints are evaluated with the raw (unsmoothed) primordial spectra.\label{fig:constraints}}
\end{figure*}

We now combine the ingredients above and map out the observationally viable parameter space of each model. For every parameter point we solve the background and linear perturbation equations numerically, compute the primordial spectrum $\mathcal{P}_\zeta(k)$ continuously from the \ac{CMB} pivot scale to the distortion scales, and evaluate the \ac{SIGW} energy density, the $\mu$ and $y$ distortions, and the \ac{PBH} abundance as described in the previous section. A point is called allowed only if it satisfies all the bounds simultaneously: the \ac{PBH} bound $f_{\mathrm{pbh}}<1$, the \ac{CMB}$+$\ac{BAO} bound of Eq.~\eqref{eq:rhup}, and the COBE/FIRAS bounds of Eq.~\eqref{eq:FIRAS}. To help identify which observation is relevant in which part of the parameter space, the dotted gray contours in Fig.~\ref{fig:constraints} mark the peak position $k_p=10^{4}$, $10^{6}$, $10^{8}\,\mathrm{Mpc}^{-1}$ of the primordial spectrum.

For the bump/dip-modified KKLT model we fix $\sigma=0.02$ and scan the $(\phi_0,A)$ plane for both signs of the feature amplitude: a bump ($A>0$) at $\phi_0\in[3.55,3.77]$ and a dip ($A<0$) at $\phi_0\in[3.80,4.20]$. To present the results compactly we introduce $\delta\equiv1-|A|/|A_{\max}(\phi_0)|$, where $|A_{\max}(\phi_0)|$ is the largest feature depth for which the inflaton still reaches the potential minimum; small $\delta$ therefore corresponds to a deep, near-critical feature and a high spectral peak. As shown in Fig.~\ref{fig:constraints}(a), the constraint patterns of the two cases are similar. For the dip case, the \ac{PBH} bound excludes $\delta\lesssim(1.3$--$4.4)\times10^{-4}$, depending on $\phi_0$, while the \ac{CMB}$+$\ac{BAO} bound is slightly weaker and excludes $\delta\lesssim(0.7$--$2.2)\times10^{-4}$. The $\mu$-distortion bound cuts into the parameter space from the right: for $\phi_0\gtrsim4.1$ the spectral peak shifts toward the distortion window, and the $\mu$ boundary rises from $\delta\simeq2.4\times10^{-4}$ at $\phi_0\simeq4.24$ to $\delta\simeq6\times10^{-3}$ at $\phi_0=4.5$. The $y$-distortion bound is always subdominant to the $\mu$ bound. For the bump case the \ac{PBH} (\ac{CMB}$+$\ac{BAO}) bound excludes $\delta\lesssim(1.3$--$1.8)\times10^{-4}$ ($\delta\lesssim(0.3$--$0.9)\times10^{-4}$), while the $\mu$ and $y$ bounds are essentially inactive in the scanned $\phi_0$ range. The remaining light-green regions are allowed; in particular the benchmark curve of Fig.~\ref{fig1} ($\phi_0=3.98$, $A=-1.933\times10^{-3}$, corresponding to $\delta\simeq4.7\times10^{-4}$) lies inside the allowed region of the dip case, with $n_s\simeq0.972$, $r\simeq0.010$, $k_p\simeq8\times10^{6}\,\mathrm{Mpc}^{-1}$, and a peak amplitude $\mathcal{P}_\zeta(k_p)\simeq2\times10^{-2}$.

For the step-modified Starobinsky model we scan the $(\phi_0,A)$ plane with $n=3$, $\sigma=10^3$, $d=0.2$ fixed, see Fig.~\ref{fig:constraints}(b). The oscillatory nature of its primordial spectra requires some care in the intermediate observables: the distortion constraints are evaluated with the raw spectra, while the \ac{SIGW} and \ac{PBH} constraints are computed from oscillation-averaged spectra. We have checked that using the raw spectra instead leaves every exclusion boundary unchanged. Explicitly, at the benchmark point $\phi_0=5$, $A=0.025$ (with $n_s\simeq0.977$, $r\simeq0.0015$) one finds $f_{\mathrm{pbh}}\sim10^{-36}$ and $\int h^2\bar{\Omega}_{\mathrm{GW},0}^{(2)}\,\mathrm{d}\ln k\simeq2\times10^{-10}$ from the raw spectrum, far below the corresponding bounds. The only exclusion in the scanned region therefore comes from the $\mu$ distortion: for $\phi_0\gtrsim5.2$ the boundary rises from $A\simeq0.022$ at $\phi_0\simeq5.21$ to $A\simeq0.035$ at $\phi_0=5.3$, while the rest of the plane is allowed.

For the double-inflection-point model we scan the $(b_3,b_5)$ plane and define $\delta\equiv|b_5|/|b_5^{\max}(b_3)|-1$, where $|b_5^{\max}(b_3)|$ denotes the analytically estimated largest value of $|b_5|$ for which the inflaton is not trapped in the ultra-slow-roll dip. In the scan, the trapped region is identified directly from the numerical background evolution; it occupies the upper-right part of Fig.~\ref{fig:constraints}(c), with a boundary descending from $\delta\gtrsim10^{-2}$ at $b_3\lesssim4.8\times10^{-4}$ to $\delta\simeq4\times10^{-4}$ at $b_3=6.5\times10^{-4}$. The \ac{PBH} bound excludes the bottom of the plane up to $\delta\simeq2.6\times10^{-5}$--$2.5\times10^{-4}$, with the boundary reaching its maximum at $b_3\simeq5.2\times10^{-4}$, and the \ac{CMB}$+$\ac{BAO} bound sits slightly below it. The $\mu$-distortion bound enters from the right at $b_3\gtrsim5.5\times10^{-4}$ and rises toward the trapped-region boundary. As a result, the allowed band between the \ac{PBH} and trapped regions narrows rapidly with increasing $b_3$ and almost closes around $b_3\simeq6\times10^{-4}$.

In summary, all three single-field inflation models possess viable parameter regions in which the primordial curvature spectrum is amplified to $\mathcal{P}_\zeta\sim10^{-2}$ around the \ac{PTA} band, while the spectral index at the \ac{CMB} scale remains within the $2\sigma$ range favored by \ac{ACT}/\ac{SPT}, and the \ac{PBH}, \ac{SIGW}, and \ac{CMB}-distortion bounds are all satisfied. Among the three, the step-modified Starobinsky model is the least restricted, with only a corner of its parameter space excluded by the $\mu$ distortion; the bump/dip-modified KKLT model retains broad allowed strips above the \ac{PBH} boundary for both signs of $A$ (further limited by the $\mu$ distortion at large $\phi_0$ in the dip case); and the double-inflection-point model has the narrowest viable window, which nearly closes at $b_3\simeq6\times10^{-4}$.

\section{PTA observations}\label{sec:PTA}

The constraints discussed in the previous section are purely cosmological, in the sense that they do not assume any particular interpretation of the \ac{PTA} signal. On the other hand, as reviewed in the Introduction, the origin of the nanohertz \ac{SGWB} reported by the \ac{PTA} collaborations remains unclear, and \acp{SIGW} sourced by enhanced small-scale curvature perturbations are among the most competitive candidates. The purpose of this section is to answer, for each of the three inflation models, the following question: can the \acp{SIGW} generated by the model dominate the current \ac{PTA} observations while the model parameters stay inside the cosmologically allowed region found in Sec.~\ref{sec:results}? To this end, we perform a Bayesian analysis of the NANOGrav 15-year data set \cite{NANOGrav:2023gor} using the \ac{SIGW} energy-density spectra computed from our models, and then confront the resulting posteriors with the constraint maps of Fig.~\ref{fig:constraints}.

We use the Hellings--Downs correlated free spectrum of the NANOGrav 15-year data set, in the form of \ac{KDE} representations of the posterior distribution of the \ac{GW} energy density at the first 14 frequency bins \cite{Lamb:2023jls,NANOGrav:2023hvm}, covering $f\simeq2\times10^{-9}$--$2.8\times10^{-8}\,\mathrm{Hz}$, which corresponds to $k\simeq1.3\times10^{6}$--$1.8\times10^{7}\,\mathrm{Mpc}^{-1}$. For a theoretical spectrum $\bar{\Omega}_{\mathrm{GW},0}^{(2)}(f;\boldsymbol{\theta})$, the log-likelihood reads
\begin{equation}
\ln\mathcal{L}(\boldsymbol{\theta})=\sum_{i=1}^{14}\ln\mathcal{L}_{i}\!\left(\bar{\Omega}_{\mathrm{GW},0}^{(2)}(f_i;\boldsymbol{\theta})\right) \ ,
\label{eq:PTAlike}
\end{equation}
where $\mathcal{L}_i$ is the \ac{KDE} likelihood of the $i$-th frequency bin. For each of the three inflation models, we evaluate $h^2\bar{\Omega}_{\mathrm{GW},0}^{(2)}(f)$ with the procedure described in Secs.~\ref{sec:models} and \ref{sec:obs} on a two-dimensional grid of the model parameters, and interpolate the spectra over the parameter space. For the KKLT potential we analyze the bump ($A>0$) and dip ($A<0$) cases separately, with $\sigma=0.02$ fixed as before, and the sampled parameters are $(\phi_0,A)$; for the double-inflection-point model the sampled parameters are $(b_3,b_5)$, and the grid covers the untrapped near-critical strip of Fig.~\ref{fig:constraints}(c). We adopt a uniform prior on $\phi_0$ (or $b_3$) over the scanned range and, at each $\phi_0$ (or $b_3$), a uniform prior on $A$ (or $b_5$) over the near-critical region in which the inflaton is not trapped by the feature of the potential, i.e. $q\equiv|A|/|A_{\max}(\phi_0)|\in[0.95,1]$ for the KKLT models, $A\in[0.015,0.051]$ for the step model, and $\delta\equiv|b_5|/|b_5^{\max}(b_3)|-1\in[8.7\times10^{-6},3.5\times10^{-4}]$ for the double-inflection-point model; the prior ranges are collected in Table~\ref{tab:PTA}. Posteriors are sampled with the dynamic nested sampler \textsc{dynesty} \cite{Speagle:2019ivv} through the \textsc{bilby} interface \cite{Ashton:2018jfp}, using 2000 live points. Following Ref.~\cite{Wu:2024qdb}, we quantify whether the \ac{SIGW} interpretation is preferred by the data through the Bayes factor $\mathcal{B}^{\mathrm{SIGW}}_{\mathrm{SMBHB}}\equiv\mathcal{Z}_{\mathrm{SIGW}}/\mathcal{Z}_{\mathrm{SMBHB}}$, the ratio of the evidence $\mathcal{Z}$ of the \ac{SIGW} fit to that of the \ac{SMBHB}-only fit to the same data with the same likelihood; $\ln\mathcal{B}^{\mathrm{SIGW}}_{\mathrm{SMBHB}}>0$ ($<0$) indicates that the data prefer (disfavor) the \ac{SIGW} interpretation relative to the \ac{SMBHB} one.

\begin{table*}[!t]
\caption{Priors, posteriors, and Bayes factors of the Bayesian analysis of the NANOGrav 15-year data set for the three inflation models, assuming that the \acp{SIGW} of each model dominate the \ac{PTA} signal.\label{tab:PTA}}
\centering
\scriptsize
\renewcommand{\arraystretch}{1.4}
\begin{tabular}{lcccc}
\hline\hline
 & \multicolumn{2}{c}{bump/dip-modified KKLT ($\sigma=0.02$)} & step-modified Starobinsky & double-inflection-point \\
\cline{2-3}
 & $A>0$ (bump) & $A<0$ (dip) & ($n=3$, $\sigma=10^3$, $d=0.2$) &  \\
\hline
prior on $\phi_0$ / $b_3$ & $[3.55,\,3.77]$ & $[3.80,\,4.20]$ & $[4.9,\,5.1]$ & $[4.4,\,6.5]\times10^{-4}$ \\
prior on $A$ / $b_5$ & $[0.95,\,1]\,A_{\max}$ & $-[0.95,\,1]\,|A_{\max}|$ & $[0.015,\,0.051]$ & $(1\!+\!\delta)b_5^{\max}$,\, $\delta\in[8.7\!\times\!10^{-6},3.5\!\times\!10^{-4}]$ \\
posterior of $\phi_0$ / $b_3$ & $3.603^{+0.042}_{-0.029}$ & $3.888^{+0.039}_{-0.040}$ & $5.036^{+0.030}_{-0.047}$ & $(4.622^{+0.041}_{-0.027})\times10^{-4}$ \\
posterior of $A$ / $b_5$ & $(1.994^{+0.032}_{-0.044})\times10^{-3}$ & $-(2.057^{+0.054}_{-0.058})\times10^{-3}$ & $(2.63^{+0.35}_{-0.12})\times10^{-2}$ & $-(7.330^{+0.101}_{-0.152})\times10^{-4}$ \\
$\ln\mathcal{B}^{\mathrm{SIGW}}_{\mathrm{SMBHB}}$ & $+0.29\pm0.11$ & $+0.98\pm0.11$ & $-24.73\pm0.10$ & $+3.08\pm0.09$ \\
\ac{PBH}-excluded fraction & $93.9\%$ & $96.3\%$ & $0\%$ & $82.3\%$ \\
dominate the \ac{PTA} signal? & excluded by \acp{PBH} & excluded by \acp{PBH} & disfavored by the data & mostly excluded by \acp{PBH} \\
\hline\hline
\end{tabular}

\vspace{0.6em}
\begin{minipage}{0.97\textwidth}
\footnotesize
\textbf{Notes.} Posteriors are quoted as medians with $68\%$ credible intervals. $\ln\mathcal{B}^{\mathrm{SIGW}}_{\mathrm{SMBHB}}$ is the natural logarithm of the Bayes factor relative to the \ac{SMBHB}-only interpretation of the same data (see the text); a positive (negative) value indicates a preference for (against) the \ac{SIGW} interpretation. For the KKLT models, the prior on $A$ at fixed $\phi_0$ is uniform on the near-critical interval shown, with $|A_{\max}(\phi_0)|$ varying over $(1.68$--$2.19)\times10^{-3}$ within the scanned $\phi_0$ ranges; for the double-inflection-point model, the prior on $b_5$ at fixed $b_3$ is uniform on the interval shown, with $|b_5^{\max}(b_3)|$ varying over $(6.50$--$14.33)\times10^{-4}$ within the scanned $b_3$ range. The \ac{PBH}-excluded fraction is the fraction of posterior samples violating the \ac{PBH} bound $f_{\mathrm{pbh}}<1$ of Sec.~\ref{sec:results}; the \ac{CMB}$+$\ac{BAO} and distortion bounds exclude no additional posterior samples in any of the four analyses. The last row states whether the model can dominate the \ac{PTA} signal within its cosmologically allowed parameter space: the \ac{SIGW} fits of the two KKLT models are as good as the \ac{SMBHB} interpretation and that of the double-inflection-point model is moderately better, but the \ac{PTA}-favored regions are almost entirely (KKLT) or mostly (double-inflection-point) excluded by the \ac{PBH} bound; for the step model the \ac{SIGW} spectrum cannot reproduce the NANOGrav data.
\end{minipage}
\end{table*}

\begin{figure*}[!t]
    \centering\includegraphics[width=1\textwidth]{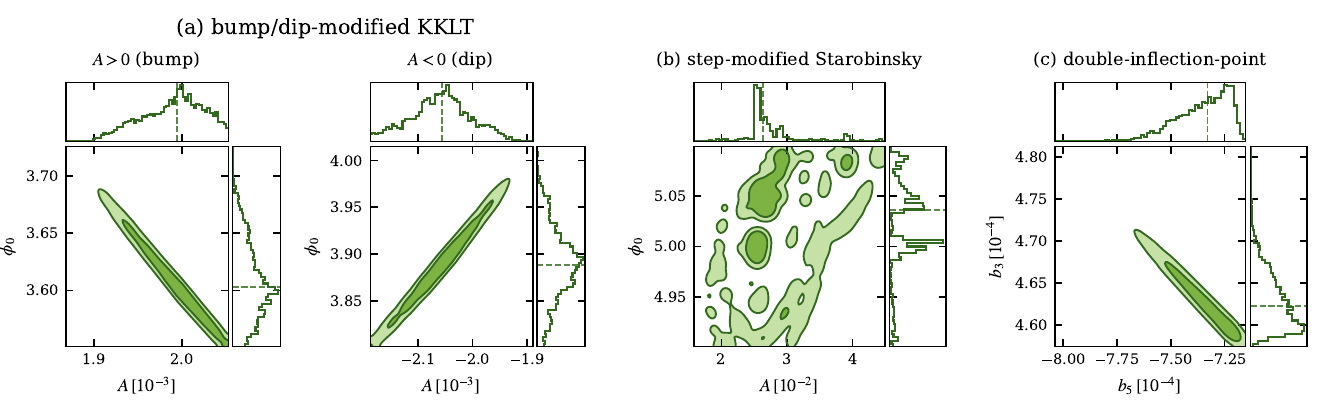}
    \caption{Posterior distributions of the model parameters from the Bayesian analysis of the NANOGrav 15-year data set, assuming that \acp{SIGW} dominate the \ac{PTA} signal. (a) The bump/dip-modified KKLT model in $(A,\phi_0)$, with the bump ($A>0$, left) and dip ($A<0$, right) cases shown separately; (b) the step-modified Starobinsky model in $(A,\phi_0)$; (c) the double-inflection-point model in $(b_5,b_3)$. The dark and light green regions denote the $68\%$ and $95\%$ credible regions, and the dashed lines mark the one-dimensional medians. The well-localized, strongly degenerate posteriors of the KKLT and double-inflection-point models contrast with the diffuse, prior-dominated posterior of the step model, the latter reflecting the poor fit of its \ac{SIGW} spectrum to the data.\label{fig:corner}}
\end{figure*}

The resulting posteriors are shown in Fig.~\ref{fig:corner} and summarized in Table~\ref{tab:PTA}. For the bump- and dip-modified KKLT models, the data select narrow, strongly degenerate regions: $A=1.994^{+0.032}_{-0.044}\times10^{-3}$ and $\phi_0=3.603^{+0.042}_{-0.029}$ for the bump case, and $A=-2.057^{+0.054}_{-0.058}\times10^{-3}$ and $\phi_0=3.888^{+0.039}_{-0.040}$ for the dip case. The Bayes factors are $\ln\mathcal{B}^{\mathrm{SIGW}}_{\mathrm{SMBHB}}=+0.29\pm0.11$ and $+0.98\pm0.11$, respectively, indicating that the \ac{SIGW} spectra of these two models account for the NANOGrav signal as well as the conventional astrophysical interpretation. In this purely spectral sense, the \acp{SIGW} of the bump/dip-modified KKLT model are able to dominate the current \ac{PTA} signal.

\begin{figure*}[!t]
    \centering\includegraphics[width=1\textwidth]{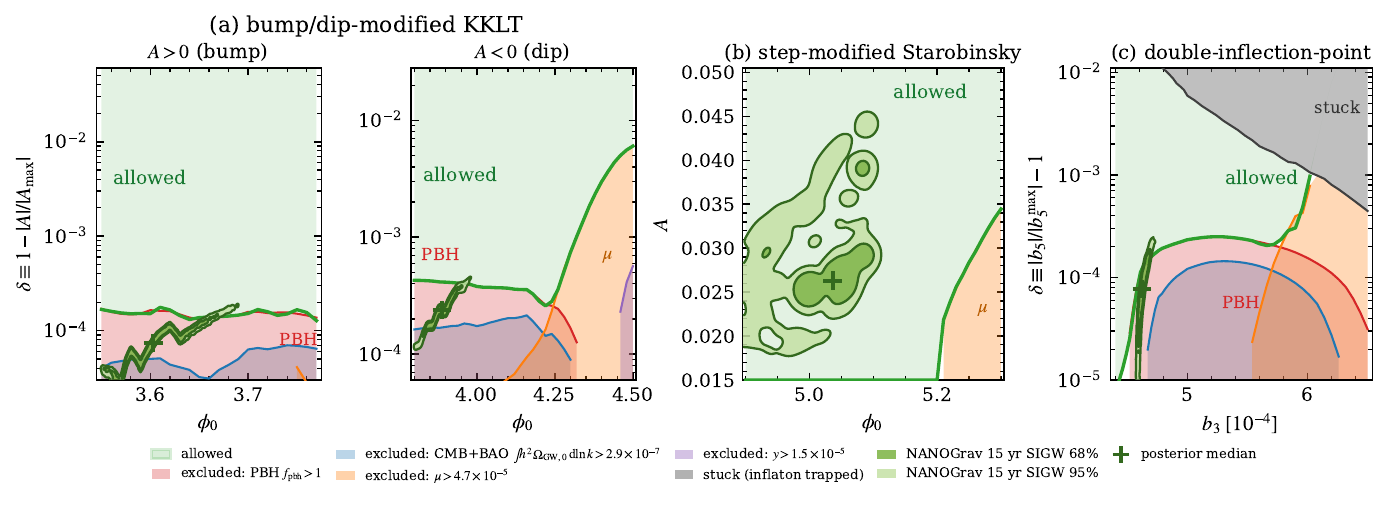}
    \caption{The $68\%$ and $95\%$ credible regions of the NANOGrav 15-year posteriors (dark and light green contours, with the posterior median marked by the cross) overlaid on the constraint maps: (a) the bump/dip-modified KKLT model in the $(\phi_0,\delta)$ plane, with the bump ($A>0$, left) and dip ($A<0$, right) cases shown separately; (b) the step-modified Starobinsky model in the $(\phi_0,A)$ plane; (c) the double-inflection-point model in the $(b_3,\delta)$ plane. The excluded regions follow the same convention as Fig.~\ref{fig:constraints}. For both KKLT cases the \ac{PTA}-favored region is pinned to the near-critical edge $\delta\rightarrow0$ and lies almost entirely inside the \ac{PBH}-excluded region. For the step model the posterior lies inside the allowed region, but the corresponding \ac{SIGW} spectra do not fit the NANOGrav data, see Table~\ref{tab:PTA}. For the double-inflection-point model the bulk of the posterior overlaps the \ac{PBH}-excluded region, while a subdominant tail at $b_3\gtrsim4.65\times10^{-4}$ survives inside the allowed band.\label{fig:overlay}}
\end{figure*}

However, the \ac{PTA}-favored regions of the KKLT models are in severe tension with the \ac{PBH} bound. This is demonstrated in Fig.~\ref{fig:overlay}, where the $68\%$ and $95\%$ credible regions of the NANOGrav posteriors are overlaid on the constraint maps: for both the bump and the dip cases the posterior is pinned to the near-critical edge $|A|\rightarrow|A_{\max}(\phi_0)|$, i.e. $\delta\rightarrow0$. The data prefer the largest available spectral peak, and hence the largest \ac{SIGW} amplitude, but this is precisely the region in which the \ac{PBH} abundance becomes excessive. Quantitatively, evaluating the \ac{PBH} abundance of each posterior sample with the pipeline of Sec.~\ref{sec:obs}, we find that $93.9\%$ ($96.3\%$) of the posterior samples of the bump (dip) model violate the bound $f_{\mathrm{pbh}}<1$; including the \ac{CMB}$+$\ac{BAO} bound does not change these fractions. Therefore, once the cosmological constraints of Sec.~\ref{sec:results} are imposed, the bump/dip-modified KKLT model cannot dominate the \ac{PTA} signal within its allowed parameter space.

For the step-modified Starobinsky model the situation is qualitatively different. The Bayes factor is $\ln\mathcal{B}^{\mathrm{SIGW}}_{\mathrm{SMBHB}}=-24.73\pm0.10$, i.e. the \ac{SIGW} interpretation is decisively disfavored compared to the \ac{SMBHB} one: the narrow and oscillatory \ac{SIGW} spectrum of this model, with its steep infrared tail, cannot reproduce the broadband shape of the NANOGrav free spectrum. Repeating the analysis with the oscillation-averaged primordial spectra yields a similar posterior and $\ln\mathcal{B}^{\mathrm{SIGW}}_{\mathrm{SMBHB}}=-26.32\pm0.10$, so this conclusion is robust. It should be emphasized that the fact that the posterior of this model lies inside the cosmologically allowed region (see Fig.~\ref{fig:overlay}) does not mean that the model can explain the \ac{PTA} signal. Since the likelihood shows no significant preference within the prior, the posterior is diffuse and essentially traces the prior itself (see Fig.~\ref{fig:corner}), and most of the prior range happens to be cosmologically allowed. Dominating the \ac{PTA} signal requires both a good spectral fit to the \ac{PTA} data and consistency with the cosmological bounds; the step model fails the former condition, and is therefore disfavored as the dominant source of the \ac{PTA} signal by the \ac{PTA} data themselves.

For the double-inflection-point model, the Bayesian analysis yields the best spectral fit among the three models. The posterior is narrow and strongly degenerate in the $(b_5,b_3)$ plane, $b_3=(4.622^{+0.041}_{-0.027})\times10^{-4}$ and $b_5=-(7.330^{+0.101}_{-0.152})\times10^{-4}$ (see Fig.~\ref{fig:corner}), and the Bayes factor $\ln\mathcal{B}^{\mathrm{SIGW}}_{\mathrm{SMBHB}}=+3.08\pm0.09$ indicates a moderate preference for the \ac{SIGW} interpretation over the \ac{SMBHB} one. This improvement over the KKLT cases is mainly due to the spectral shape: the peak position of the primordial spectrum shifts continuously with $b_3$ (see the dotted $k_p$ contours in Fig.~\ref{fig:constraints}(c)), so that the broad \ac{SIGW} peak can be placed directly inside the most sensitive frequency bins of NANOGrav. However, the \ac{PTA}-favored region is again in strong tension with the \ac{PBH} bound. Evaluating the \ac{PBH} abundance of each posterior sample with the pipeline of Sec.~\ref{sec:obs}, we find that $82.3\%$ of the samples violate the bound $f_{\mathrm{pbh}}<1$, while the \ac{CMB}$+$\ac{BAO} and distortion bounds exclude no additional samples. As shown in Fig.~\ref{fig:overlay}(c), only a subdominant posterior tail at $b_3\gtrsim4.65\times10^{-4}$ survives inside the allowed band, where the spectral peak moves toward the high-frequency edge of the \ac{PTA} band and a smaller peak amplitude suffices to fit the data. The intersection between the \ac{PTA}-favored region and the cosmologically allowed window is therefore non-empty, but it does not cover the bulk of the posterior: like the bump/dip-modified KKLT model, the double-inflection-point model cannot dominate the \ac{PTA} signal within its allowed parameter space.

The preceding analysis demonstrates that the two single-field inflation models considered in cases (a) and case (c)  can reproduce the ACT/SPT measurements while providing a SIGW-dominated explanation of the PTA signal. Nevertheless, this interpretation is in tension with the observational upper limit on the PBH abundance. Possible ways to alleviate this tension include incorporating primordial non-Gaussianity \cite{Franciolini:2023pbf,Liu:2023ymk,Wang:2023ost,Choudhury:2023fwk,Choudhury:2024kjj,Choudhury:2024dzw}, allowing the sound-speed parameter to vary \cite{Balaji:2023ehk,Chen:2024fir,Liu:2023hpw}, and accounting for third-order SIGWs \cite{Wang:2023sij,Zhao:2026jax,Zhou:2025djn}. We clarify that the constraints on single-field inflation models imposed by current cosmological observations are those derived in Sec.~\ref{sec:results}. The results presented in this section should instead be understood as conditional results obtained under the assumption that SIGWs dominate the PTA observations.

\section{Conclusion}\label{sec:conclusion}

In this work, we have investigated three classes of single-field inflation models that can generate large primordial curvature perturbations at sub-Mpc scales while predicting a scalar spectral index $n_s$ consistent with the \ac{ACT}/\ac{SPT} measurements: the bump/dip-modified KKLT potential, the step-modified Starobinsky potential, and the double-inflection-point potential. For each model we solved the background and linear perturbation equations numerically, obtained the primordial power spectrum $\mathcal{P}_{\zeta}(k)$ continuously from the \ac{CMB} pivot scale to the distortion scales, and computed the resulting \ac{SIGW} energy-density spectrum, the \ac{CMB} $\mu$- and $y$-type spectral distortions, and the \ac{PBH} abundance.

Scanning the parameter spaces, we mapped out the regions that simultaneously satisfy the \ac{PBH} bound $f_{\mathrm{pbh}}<1$, the \ac{CMB}$+$\ac{BAO} bound on the total \ac{SIGW} energy density, and the COBE/FIRAS distortion bounds, as shown in Fig.~\ref{fig:constraints}. All three models possess allowed regions in which the primordial spectrum is amplified to $\mathcal{P}_{\zeta}\sim10^{-2}$ in the \ac{PTA} band while $n_s$ remains within the $2\sigma$ range favored by \ac{ACT}/\ac{SPT}. Among them, the step-modified Starobinsky model is the least restricted, with only a corner of its parameter space excluded by the $\mu$ distortion; the bump/dip-modified KKLT model retains broad allowed strips above the \ac{PBH} boundary for both signs of $A$; and the double-inflection-point model has the narrowest viable window, which nearly closes at $b_3\simeq6\times10^{-4}$.

We then examined whether the \acp{SIGW} produced by these models can dominate the nanohertz \ac{SGWB} observed by the \ac{PTA} collaborations. Our Bayesian analyses of the NANOGrav 15-year data set show that the answer is essentially negative once the cosmological constraints are taken into account. For the bump- and dip-modified KKLT models, the \ac{SIGW} spectra fit the data as well as the conventional \ac{SMBHB} interpretation, but the \ac{PTA}-favored region is pinned to the near-critical edge of the parameter space, and $93.9\%$ ($96.3\%$) of the posterior samples of the bump (dip) model are excluded by the \ac{PBH} bound. For the step-modified Starobinsky model, the \ac{SIGW} spectrum cannot reproduce the broadband shape of the NANOGrav data, with a Bayes factor $\ln\mathcal{B}^{\mathrm{SIGW}}_{\mathrm{SMBHB}}\simeq-25$ relative to the \ac{SMBHB} interpretation. The double-inflection-point model provides the best spectral fit, with $\ln\mathcal{B}^{\mathrm{SIGW}}_{\mathrm{SMBHB}}\simeq+3.1$ moderately favoring the \ac{SIGW} interpretation, since its spectral peak can be placed directly inside the most sensitive \ac{PTA} bins; nevertheless, $82.3\%$ of its posterior samples are again excluded by the \ac{PBH} bound, and only a subdominant posterior tail at $b_3\gtrsim4.65\times10^{-4}$ survives inside the allowed window. In all three models, therefore, the region preferred by the \ac{PTA} data is in strong tension with the cosmological bounds, and none of them can dominate the current \ac{PTA} signal within its allowed parameter space.

Our results demonstrate that combining observations across vastly different scales, including the \ac{CMB} anisotropies and spectral distortions, \ac{BAO}, the \ac{PBH} abundance, and \ac{PTA} measurements, provides a powerful and necessary tool to discriminate among single-field inflation models with enhanced small-scale perturbations: a model that fits the \ac{ACT}/\ac{SPT} spectral index and produces a sizable \ac{SIGW} signal may still be ruled out by the \ac{PBH} overproduction bound or by the spectral shape favored by the \ac{PTA} data. Future \ac{PTA} observations with higher sensitivity, next-generation \ac{CMB} spectral-distortion experiments, and space-based \ac{GW} detectors such as \ac{LISA} are expected to further shrink the viable parameter space of these models.

\section*{Data availability}\label{sec:6.0}
The data that support the findings of this article are openly available in Ref. \cite{Nanograv:KDE}.

\vspace{0.3cm}
\begin{acknowledgements} 
This work has been funded by the National Nature Science Foundation of China under grant No. 12447127. We thank Dr. Di Wu\orcidlink{0000-0001-7309-574X} for providing the scripts used for the PTA data analysis.
\end{acknowledgements}

\bibliography{ref.bib}

\end{document}